\documentclass[sigconf]{acmart}

\usepackage{threeparttable}
\usepackage{multirow}
\usepackage{algorithm}
\usepackage{algpseudocode}
\usepackage{graphicx}
\usepackage{color}
\usepackage{siunitx}
\usepackage{stfloats}
\usepackage{enumitem}

\begin{document}

\title{HOLMES: In-Context Failure-Center Localization for High-Dimensional Yield Estimation}



\author{
Wei W. Xing$^{1}$,
Xixi Zhou$^2$,
Kaiqi Huang$^2$,
Jiaye Pan$^2$,
Hong Qiu$^2$,
Xin Wang$^2$,
Shan Shen$^{3,*}$
}

\affiliation{
  \institution{$^1$University of Sheffield, Sheffield, UK}
  \institution{$^2$SZU–UoS Joint Centre for Innovation and Entrepreneurship, Shenzhen University, Shenzhen, China}
  \institution{$^3$Nanjing University of Science and Technology, Nanjing, China}
  \country{}
}

\email{w.xing@sheffield.ac.uk, shanshen@njust.edu.cn}

\thanks{
This work was supported by the Fundamental Research Funds for the Central Universities (Grant No. 30925010605, 30924012004), Jiangsu Provincial Major Science and Technology Project (Grant No. BG2025012).
$^*$Corresponding author: S. Shen.
}

\begin{abstract}
Importance sampling for high-sigma yield estimation requires locating the failure center from a severely imbalanced sample set.  Existing surrogate-assisted methods rely on iterative gradient-based training, ill-posed under extreme class imbalance; model errors propagate into the estimator, causing accuracy collapse in high dimensions.  We recast failure-center localization as few-shot binary classification: a prior-fitted tabular foundation model performs gradient-free in-context inference in a single forward pass, eliminating the ill-posed training loop.  \textbf{HOLMES} (High-sigma Optimal Localization via Manifold Estimation and Sampling) pairs this with an SVD-based anisotropic proposal that captures the local geometry of the failure manifold, and a hit-rate-driven adaptive mixing scheme that stabilizes importance weights where conventional adaptation collapses.  On 6T SRAM benchmarks spanning $D = 108$ to $D = 1{,}152$, full-dimensional baselines exhibit accuracy collapse at some dimension, with the strongest baseline reaching 25.8\% relative error; PCA+MNIS is additionally evaluated at the two largest dimensions.  HOLMES remains within 5.9\% across all five configurations with up to $58.8\times$ speedup over Monte Carlo. The code is available on \href{https://github.com/IceLab-JCIE/ICE006-Yield-Holmes}{GitHub}.

\end{abstract}

\settopmatter{printacmref=false}
\pagestyle{plain}

\copyrightyear{2026}
\acmYear{2026}
\setcopyright{cc}
\setcctype{by}
\acmConference[ICCAD '26]{IEEE/ACM International Conference on Computer-Aided Design}{November 08--12, 2026}{San Jose, CA, USA}
\acmBooktitle{IEEE/ACM International Conference on Computer-Aided Design (ICCAD '26), November 08--12, 2026, San Jose, CA, USA}
\acmDOI{10.1145/3831252.3834224}
\acmISBN{979-8-4007-2873-0/2026/11}

\maketitle

\section{Introduction}
\label{sec:intro}

As SRAM arrays scale to advanced technology nodes, process variations such as intra-die mismatches, doping fluctuations, and threshold voltage shifts drive circuit performance outside specification, making yield estimation a critical bottleneck in memory design~\cite{openyield}.
The severity of this challenge grows with array size: a $3\times 2$ array introduces $D=108$ variation parameters, while an $8\times 8$ array reaches $D=1{,}152$, each representing an independent source of random variation.
The reference estimator, Monte Carlo (MC) simulation, requires $\mathcal{O}(1/P_f)$ SPICE calls, demanding $10^7$ or more evaluations for the high-sigma targets of modern SRAM arrays~\cite{mnis}.
Importance sampling (IS) reduces this cost by concentrating samples near the failure region, but locating that region grows difficult as $D$ scales to hundreds of dimensions.

Two lines of work have addressed this problem.
The first refines the IS proposal distribution directly, progressing from norm minimization to locate the most probable failure point~\cite{mnis}, through adaptive updating~\cite{ais} and multi-region coverage~\cite{hscs,acs}, to variational frameworks that derive closed-form optimal proposals~\cite{vis}.
The second trains surrogate models, including Gaussian processes, deep kernel classifiers, and normalizing flows, to approximate the circuit response or failure boundary, substituting cheap model evaluations for expensive SPICE calls~\cite{lrta,bya,asdk,fusis}.
Efforts to unify both lines have produced frameworks that jointly optimize proposals and surrogates~\cite{opt,optimis,YMCA}, yet their performance degrades sharply as dimensionality grows.

Despite their architectural differences, all existing methods share a common bottleneck: locating the failure center, the highest-density point of the failure region under the prior, from an inherently imbalanced sample set.
Surrogate-assisted IS methods address this through iterative gradient-based training, which is ill-posed under extreme class imbalance; model errors propagate directly into the IS estimator, with reported failure rates exceeding five out of ten runs in high-dimensional settings~\cite{optimis}.
Analytic methods avoid gradient training by imposing strong distributional assumptions that enable closed-form solutions from failure samples~\cite{shen2022timing}; these assumptions limit expressiveness when the failure manifold deviates from the assumed parametric form.
Even methods that focus on proposal optimality alone are limited by localization quality: as $D$ scales from 576 to $1{,}152$, the relative estimation error of MNIS~\cite{mnis}, the strongest parametric baseline, rises from ${\sim}5\%$ to 67.8\%, not because its proposal is poorly shaped, but because the failure-center estimate degrades as failures become sparse.

All three limitations trace to the same miscast formulation: locating the failure center has been treated as a regression or parametric fitting problem.
IS requires only identifying the \emph{highest-density point} of the failure manifold from a labeled set, a \emph{few-shot binary classification} problem.
Classification requires only a decision boundary between fail and pass regions, not a global model of circuit response, and is inherently more robust to the extreme class imbalance that destabilizes regression and density estimation.
Recognizing this reformulation enables gradient-free in-context inference via a prior-fitted tabular foundation model~\cite{tabpfn}, which performs Bayesian classification over a labeled context in a single forward pass, eliminating the ill-posed training loop entirely.

\begin{figure*}[!t]
    \centering
    \includegraphics[width=\textwidth]{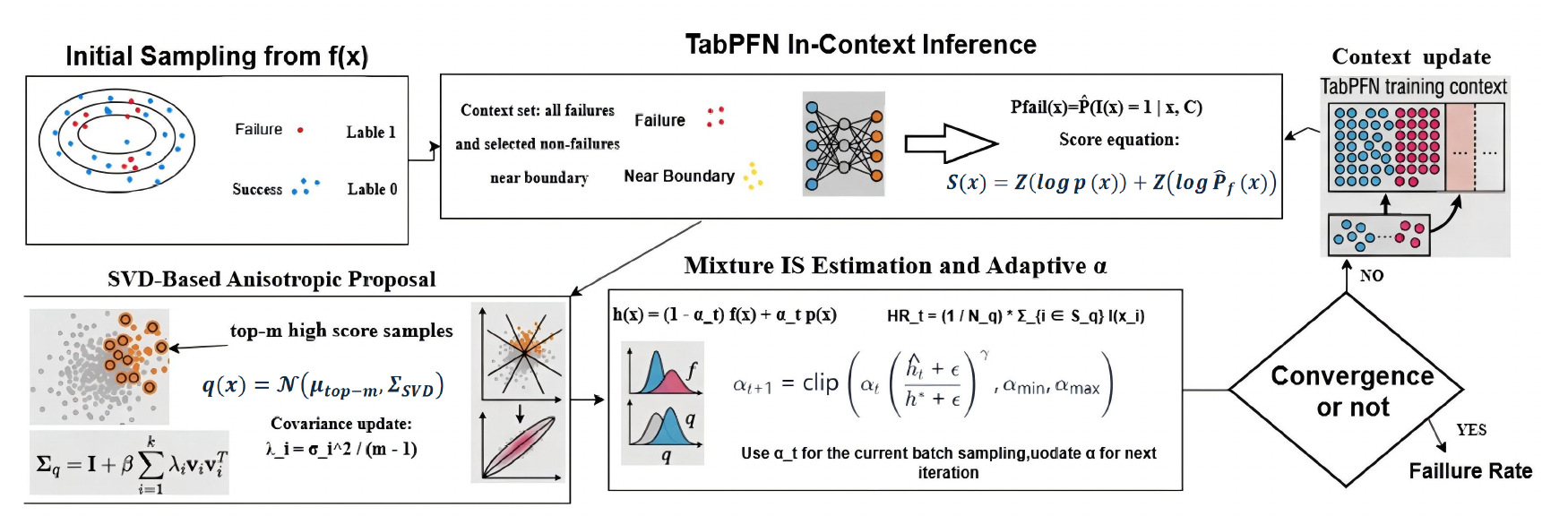}
    \vspace{-10pt}
    \caption{Overview of HOLMES. (1)~Initial samples are drawn from the prior $p(\mathbf{x})$ and evaluated via SPICE to obtain failure labels. (2)~A failure-preserving context set $\mathcal{C}$ is constructed and fed to TabPFN, which outputs failure probabilities via in-context inference without gradient updates; the separately z-normalized log-prior and log-failure-probability terms are summed to rank the candidates. (3)~The mean of the top-$m$ high-score samples defines the proposal center $\boldsymbol{\mu}_q$, and SVD of the samples centered at $\boldsymbol{\mu}_q$ yields an anisotropic proposal $q(\mathbf{x})$ aligned with the failure manifold. (4)~Samples are drawn from the mixture $h(\mathbf{x}) = (1-\alpha_t)p(\mathbf{x}) + \alpha_t q(\mathbf{x})$, and $\alpha_t$ is adapted via the hit rate. (5)~The process repeats until the coefficient of variation of $\hat{P}_f$ falls below the convergence threshold.}
    \label{fig:holmes_illustration}
\end{figure*}

\textbf{HOLMES} (High-sigma Optimal Localization via Manifold Estimation and Sampling) instantiates this insight with three components: a failure-preserving context construction that retains failure samples under extreme imbalance, an SVD-based anisotropic proposal that aligns sampling with the principal directions of the failure manifold, and a hit-rate-driven adaptive mixing scheme that replaces effective-sample-size (ESS)-based adaptation, which collapses in high dimensions.
On 6T SRAM benchmarks spanning $D = 108$ to $D = 1{,}152$, full-dimensional baselines exhibit accuracy collapse at some dimension, with the strongest baseline reaching 25.8\% relative error; PCA+MNIS is additionally evaluated at $D=864$ and $D=1152$. HOLMES remains within 5.9\% across all five configurations while requiring only $1{,}700$ to $13{,}000$ SPICE simulations.
Figure~\ref{fig:holmes_illustration} illustrates the complete workflow.
The contributions of this work are:
\begin{itemize}[nosep,leftmargin=*]
    \item We identify failure-center localization as a few-shot tabular classification problem and show that this reformulation enables gradient-free in-context inference via a prior-fitted tabular foundation model, eliminating the ill-posed training loop shared by all surrogate-assisted IS methods.
    \item We introduce a failure-preserving context construction that ensures failure samples are retained in the in-context set under extreme class imbalance, enabling reliable localization from as few as one or two failure instances per iteration.
    \item An SVD-based anisotropic IS proposal adapts to the local geometry of the failure manifold; removing this component increases relative error from 0.1\% to 11.88\% at $D = 864$.
    \item A hit-rate-driven adaptive mixing scheme stabilizes importance weights where ESS-based adaptation degrades; removing this component increases relative error from 0.1\% to 5.65\% at $D = 864$.
\end{itemize}

\section{Preliminaries}
\label{sec:background}
\subsection{Problem Definition}

Let $\mathbf{x} = [x^{(1)}, \ldots, x^{(D)}]^T \in \mathbb{R}^D$ denote the process variation parameters, assumed independent and standard Gaussian: $p(\mathbf{x}) = \mathcal{N}(\mathbf{0}, \mathbf{I})$.
Given $\mathbf{x}$, a SPICE simulator evaluates circuit performance $y = f(\mathbf{x})$.
A design fails if $y$ violates a predefined specification; the failure indicator is $I(\mathbf{x}) \in \{0, 1\}$, and the failure rate is
\begin{equation}
    P_f = \int_{\mathbb{R}^D} I(\mathbf{x})\, p(\mathbf{x})\, d\mathbf{x}.
\end{equation}
The goal is to estimate $P_f$ accurately with as few SPICE calls as possible.

\subsection{Monte Carlo and Importance Sampling}

Monte Carlo (MC) estimation draws $N$ samples from $p(\mathbf{x})$ and computes $\hat{P}_f = \frac{1}{N}\sum_{i=1}^{N} I(\mathbf{x}_i)$.
To achieve $90\%$ accuracy with $90\%$ confidence, $N \approx 100 / P_f$ samples are required~\cite{mnis}, demanding $10^7$ or more SPICE evaluations for high-sigma targets.

Importance sampling (IS) reduces this cost by drawing samples from a proposal distribution $q(\mathbf{x})$ and reweighting:
\begin{equation}
    \hat{P}_f = \frac{1}{N} \sum_{i=1}^{N} I(\mathbf{x}_i)\frac{p(\mathbf{x}_i)}{q(\mathbf{x}_i)}, \quad \mathbf{x}_i \sim q(\mathbf{x}).
\end{equation}
The variance of this estimator is minimized by the optimal $q^*(\mathbf{x})$~\cite{opt}
\begin{equation}
    q^*(\mathbf{x}) = \frac{I(\mathbf{x})\,p(\mathbf{x})}{P_f},
    \label{eq:optimal_q}
\end{equation}
which concentrates mass precisely over the failure region, weighted by prior density.

\subsection{The Failure Center}
\label{sec:background_fc}

Equation~\eqref{eq:optimal_q} is intractable in practice, but its mode reveals a tractable surrogate objective.
The highest-density point of $q^*(\mathbf{x})$, the \emph{failure center}, is
\begin{equation}
    \mathbf{x}^* = \underset{\mathbf{x}:\, I(\mathbf{x})=1}{\arg\max}\ p(\mathbf{x}),
    \label{eq:failure_center}
\end{equation}
the failure-region point with maximum prior probability.
Since $p(\mathbf{x}) = \mathcal{N}(\mathbf{0}, \mathbf{I})$ decreases monotonically with $\|\mathbf{x}\|$, Eq.~\eqref{eq:failure_center} reduces to the classic minimum-norm failure point targeted by MNIS~\cite{mnis} and subsequent methods~\cite{vis,opt}.
Centering the IS proposal near $\mathbf{x}^*$ reduces estimator variance: the closer the proposal center is to $\mathbf{x}^*$, the fewer samples are wasted outside the failure region.

In practice, $\mathbf{x}^*$ must be located from a labeled dataset $\{(\mathbf{x}_i, I(\mathbf{x}_i))\}$ accumulated across IS iterations, in which each iteration yields at most a handful of failures against thousands of non-failures.
Locating $\mathbf{x}^*$ from this severely imbalanced set is a \emph{few-shot binary classification} problem: inferring which candidate points lie in the failure region from a small labeled context.

\subsection{Prior-Fitted Networks for Tabular Classification}
\label{sec:background_tabpfn}

TabPFN~\cite{tabpfn} is a tabular foundation model that performs Bayesian classification via \emph{in-context learning}: given a labeled context set $\mathcal{C} = \{(\mathbf{x}_i, I(\mathbf{x}_i))\}_{i=1}^n$, it outputs the posterior predictive distribution $\hat{P}(I(\mathbf{x})=1 \mid \mathbf{x}, \mathcal{C})$ for any query point $\mathbf{x}$ in a single forward pass, without gradient updates.
TabPFN is pre-trained on synthetic datasets drawn from a prior over structural causal models, approximating Bayesian inference over this prior at inference time.

Three properties make TabPFN suited to failure-center localization.
First, inference requires no parameter updates: the model adapts to a new context set solely through attention, making it immune to training instability under extreme class imbalance.
Second, the model is calibrated for small sample sizes by construction, as its prior was learned from datasets of scale comparable to the context sets encountered in IS iterations.
Third, inference produces posterior probabilities in a single forward pass, enabling rapid re-estimation as new samples accumulate across IS iterations.
TabPFN imposes two practical constraints: a maximum input dimensionality $D_{\max}$ and a bounded context size $C_{\max}$, both of which require adaptation for high-dimensional yield estimation.

\section{Proposed Method}
\label{sec:method}

\subsection{Task Reformulation}
\label{sec:reformulation}

Given a labeled dataset $\mathcal{D} = \{(\mathbf{x}_i, I(\mathbf{x}_i))\}_{i=1}^N$ accumulated across IS iterations, we score each sample in $\mathcal{A}=\mathcal{D}_f\cup\mathcal{X}_{\text{cand}}$ as
\begin{equation}
    s(\mathbf{x}) = \mathcal{Z}_{\mathcal{A}}[\log p(\mathbf{x})]
    + \mathcal{Z}_{\mathcal{A}}[\log \hat{P}(I(\mathbf{x})=1 \mid \mathbf{x}, \mathcal{C})],
    \label{eq:score}
\end{equation}
where $\mathcal{D}_f \subset \mathcal{D}$ is the set of observed failure samples, $\mathcal{X}_{\text{cand}}$ is a candidate pool drawn from the current proposal, $\mathcal{C}\subset\mathcal{D}$ is the classifier context, and $\mathcal{Z}_{\mathcal{A}}[\cdot]$ denotes z-normalization over $\mathcal{A}$. Thus, the log-prior and predicted log-failure-probability components are normalized separately before addition, preventing their different numerical scales from dominating the ranking. Using the soft probability $\hat{P}$ rather than the hard indicator $I(\mathbf{x})$ extends the search beyond observed failures and ranks unvisited candidates by predicted failure likelihood.
At the first iteration, no proposal exists; $\mathcal{X}_{\text{cand}}$ is drawn from $p(\mathbf{x})$ and $\mathcal{C}$ consists of the initial failure samples paired with uniformly drawn non-failures.

Equation~\eqref{eq:score} places concrete demands on the classifier.
The context set contains as few as one or two failure instances per iteration, so any gradient-based classifier is ill-posed under this imbalance regardless of regularization.
Calibrated estimates are required across iterations without retraining, since online gradient updates risk propagating early errors forward.
The classifier must also generalize from a handful of labeled examples without memorizing the rare failure class.
TabPFN~\cite{tabpfn} satisfies all three demands by construction: rather than optimizing parameters on $\mathcal{D}$, it treats $\mathcal{D}$ as a context and performs approximate Bayesian inference via a single transformer forward pass, requiring no gradient computation and no retraining between iterations.
Its pre-trained prior biases toward low-complexity decision boundaries, consistent with the local structure of the failure manifold near $\mathbf{x}^*$, while remaining flexible enough to capture nonlinear boundaries when evidence supports it.
The two constraints identified in Section~\ref{sec:background_tabpfn} are addressed by feature selection (Section~\ref{sec:feature}) and the context construction strategy described next.

\subsection{In-Context Failure-Center Localization}
\label{sec:localization}

The quality of in-context inference depends critically on the composition of the context set $\mathcal{C}$.
Under extreme class imbalance, randomly subsampling $\mathcal{D}$ to fit the context budget yields sets in which failure samples are absent or represented by a single instance, preventing the model from locating the failure boundary.
We address this with a \emph{failure-preserving} construction: failure samples are allocated budget first, with non-failures filling the remainder.

Let $\mathcal{F}$ and $\mathcal{N}$ denote the failure and non-failure subsets of $\mathcal{D}$, and let $C_{\max}$ be the context budget.
Failure samples are selected as
\begin{equation}
    |\mathcal{F}_{\text{sel}}| = \min\!\left(|\mathcal{F}|,\ \tfrac{C_{\max}}{2}\right),
\end{equation}
retaining all failures when they are scarce while preventing their domination when they are abundant.
Non-failures fill the remaining capacity, $|\mathcal{N}_{\text{sel}}| = C_{\max} - |\mathcal{F}_{\text{sel}}|$, selected to represent the decision boundary region.
The context set is $\mathcal{C} = \mathcal{F}_{\text{sel}} \cup \mathcal{N}_{\text{sel}}$.

Given $\mathcal{C}$, TabPFN outputs $\hat{P}(I(\mathbf{x})=1 \mid \mathbf{x}, \mathcal{C})$ for all candidates in a single forward pass. We then evaluate the standardized score in Eq.~\eqref{eq:score} over $\mathcal{A}=\mathcal{D}_f \cup \mathcal{X}_{\text{cand}}$ and retain its top-$m$ samples as the high-risk neighborhood used to construct the proposal.

\subsection{Feature Selection for High-Dimensional Inputs}
\label{sec:feature}

When $D$ exceeds TabPFN's input limit $D_{\max}$, we retain the $K = D_{\max}$ features most informative of failure.
Because samples in $\mathcal{D}$ are drawn from the mixture distribution $h(\mathbf{x})$ rather than $p(\mathbf{x})$, direct mutual information (MI) estimation is biased toward regions overrepresented by the proposal.
Without correction, features informative under $h(\mathbf{x})$ would be favored over those informative under $p(\mathbf{x})$, distorting the failure center estimate toward the current proposal rather than the true $\mathbf{x}^*$.
We correct this via importance weighting: each sample $\mathbf{x}_i$ receives weight
\begin{equation}
    \tilde{w}_i \propto \frac{p(\mathbf{x}_i)}{h(\mathbf{x}_i)},
\end{equation}
and a resampled dataset approximating $p(\mathbf{x})$ is constructed by drawing according to $\tilde{w}_i$.
MI between each feature and the failure indicator is then estimated on the resampled data:
\begin{equation}
    \mathrm{MI}(x^{(j)}; I) = \sum_{x^{(j)},\, I} p(x^{(j)}, I)\log \frac{p(x^{(j)}, I)}{p(x^{(j)})\,p(I)},
\end{equation}
and the top-$K$ features are retained:
\begin{equation}
    \mathcal{S} = \operatorname{TopK}\!\left(\{\mathrm{MI}(x^{(j)}; I)\}_{j=1}^{D},\ K\right).
\end{equation}
The reduced feature vector $\mathbf{x}_{\mathcal{S}}$ is used as input to TabPFN.

\subsection{SVD-Based Anisotropic Proposal Construction}
\label{sec:proposal}

Locating the high-risk region is not sufficient: the proposal must also capture the local geometry of the failure manifold to concentrate samples efficiently. An isotropic Gaussian wastes sampling budget on the many directions orthogonal to the manifold; replacing the anisotropic proposal with an isotropic one increases relative error from 0.1\% to 11.88\% at $D = 864$ (Table~\ref{tab:ablation}).

Let $\mathcal{H}$ denote the top-$m$ samples in $\mathcal{A}$ ranked by score $s(\mathbf{x})$. Their mean defines the proposal center:
\begin{equation}
    \boldsymbol{\mu}_q = \frac{1}{m}\sum_{\mathbf{x}\in\mathcal{H}}\mathbf{x}.
    \label{eq:proposal_center}
\end{equation}
Here, $\mathbf{x}^*$ denotes the theoretical failure center, whereas $\boldsymbol{\mu}_q$ is the empirical proposal center obtained from the top-$m$ high-score samples.
The same top-$m$ mean is used to center the samples: the row associated with $\mathbf{x}_i\in\mathcal{H}$ in $X_c\in\mathbb{R}^{m\times D}$ is $(\mathbf{x}_i-\boldsymbol{\mu}_q)^T$. We then compute the singular value decomposition $X_c=U\Sigma V^T$.
The right singular vectors $\{\mathbf{v}_i\}$ identify the principal directions of variation in the high-risk neighborhood; the corresponding singular values $\{\sigma_i\}$ quantify the extent of that variation.
The proposal covariance is
\begin{equation}
    \boldsymbol{\Sigma}_q = \mathbf{I} + \beta \sum_{i=1}^{r} \lambda_i \mathbf{v}_i \mathbf{v}_i^T, \quad \lambda_i = \frac{\sigma_i^2}{m-1},
    \label{eq:sigma}
\end{equation}
where $r$ is the number of retained singular components and $\beta$ is a scaling factor.
The proposal distribution is
\begin{equation}
    q(\mathbf{x}) = \mathcal{N}(\boldsymbol{\mu}_q, \boldsymbol{\Sigma}_q),
\end{equation}
which stretches the sampling ellipsoid along the principal failure directions without requiring explicit knowledge of the failure boundary geometry.
This construction exploits a geometric property that strengthens with dimensionality: as $D$ grows, the failure manifold occupies an increasingly thin subspace of the ambient variation space, making the leading singular directions more distinct from the isotropic background and the anisotropic proposal correspondingly more efficient.

\subsection{Hit-Rate-Driven Adaptive Mixing}
\label{sec:adaptive}

Centering the proposal at the top-$m$ mean $\boldsymbol{\mu}_q$ improves exploitation, but pure exploitation risks missing failure regions not yet captured by the current estimate.
We balance this via the mixture distribution
\begin{equation}
    h(\mathbf{x}) = (1 - \alpha_t)\,p(\mathbf{x}) + \alpha_t\,q(\mathbf{x}),
\end{equation}
from which samples are drawn and the failure rate estimated as
\begin{equation}
    \hat{P}_f = \frac{1}{N} \sum_{i=1}^{N} I(\mathbf{x}_i)\frac{p(\mathbf{x}_i)}{h(\mathbf{x}_i)}, \quad \mathbf{x}_i \sim h(\mathbf{x}).
\end{equation}

The mixture coefficient $\alpha_t$ is adapted based on how effectively $q(\mathbf{x})$ discovers failures.
The conventional adaptation signal, ESS, is driven by the KL divergence between $q$ and $p$, which grows with dimension and causes ESS to collapse even when $q$ is well-positioned~\cite{vis}; in high-dimensional settings, ESS conflates distributional distance with proposal quality, making it an unreliable signal.
We instead adapt $\alpha_t$ using the \emph{hit rate}, the fraction of proposal samples that are failures:
\begin{equation}
    \mathrm{HR}_t = \frac{1}{N_q} \sum_{i \in \mathcal{I}_q} I(\mathbf{x}_i),
\end{equation}
where $N_q$ is the number of samples drawn from $q(\mathbf{x})$ at iteration $t$ and $\mathcal{I}_q$ is their index set.
The update rule is
\begin{equation}
    \alpha_{t+1} = \mathrm{clip}\!\left(\alpha_t \cdot \Bigl(\frac{\mathrm{HR}_t + \epsilon}{\tau + \epsilon}\Bigr)^{\!\gamma},\ [\alpha_{\min}, \alpha_{\max}]\right),
    \label{eq:alpha_update}
\end{equation}
where $\tau$ is the target hit rate, $\epsilon > 0$ is a smoothing constant, $\gamma \in (0,1]$ is a damping factor, and the clip operation enforces stability bounds.
When $\mathrm{HR}_t > \tau$, the proposal is effective and $\alpha_t$ increases; when $\mathrm{HR}_t < \tau$, $\alpha_t$ decreases to restore exploration of $p(\mathbf{x})$.
When $N_q$ falls below a minimum threshold, $\alpha_t$ is held constant to prevent premature collapse from unreliable hit rate estimates.
Convergence is assessed via the coefficient of variation of the failure rate estimate over the last $M$ iterations,
$$\mathrm{FOM}^{(t)} = \mathrm{Std}(\hat{P}_f^{(t-M+1:t)}) / \mathrm{Mean}(\hat{P}_f^{(t-M+1:t)});$$
sampling terminates when $\mathrm{FOM}^{(t)} < \varepsilon$ and $t \geq T_{\min}$.
The complete procedure is summarized in Algorithm~\ref{alg:holmes}.

\begin{algorithm}[t]
\small
\caption{HOLMES: High-sigma Optimal Localization via Manifold Estimation and Sampling}
\label{alg:holmes}
\begin{algorithmic}[1]
\Require Initial dataset $\mathcal{D}_0$ sampled from $p(\mathbf{x})$, maximum iterations $T$, minimum iterations $T_{\min}$, context budget $C_{\max}$, initial mixture coefficient $\alpha_0$
\Ensure  Estimated failure rate $\hat{P}_f$
\State $\mathcal{D} \leftarrow \mathcal{D}_0$, $q \leftarrow p$ \Comment{No proposal at initialization}
\For{$t = 1$ to $T$}
    \If{$D > D_{\max}$}
        \State Select top-$K$ features via importance-weighted MI (Sec.~\ref{sec:feature})
    \EndIf
    \State Construct failure-preserving context set $\mathcal{C}$ from $\mathcal{D}$ (Sec.~\ref{sec:localization})
    \State Compute $\hat{P}(I(\mathbf{x})=1 \mid \mathbf{x}, \mathcal{C})$ via TabPFN for all candidates
    \State Compute the z-normalized score $s(\mathbf{x})$ via Eq.~\eqref{eq:score}
    \State Select the top-$m$ set $\mathcal{H}$ and compute $\boldsymbol{\mu}_q$ via Eq.~\eqref{eq:proposal_center}
    \State Center $\mathcal{H}$ at $\boldsymbol{\mu}_q$ and construct $\boldsymbol{\Sigma}_q$ via SVD (Sec.~\ref{sec:proposal})
    \State Set $q(\mathbf{x}) \leftarrow \mathcal{N}(\boldsymbol{\mu}_q, \boldsymbol{\Sigma}_q)$
    \State Sample $\{\mathbf{x}_i\}$ from $h_t = (1-\alpha_t)p + \alpha_t q$
    \State Evaluate $I(\mathbf{x}_i)$ via SPICE
    \State $\mathcal{D} \leftarrow \mathcal{D} \cup \{(\mathbf{x}_i, I(\mathbf{x}_i))\}$
    \State Update $\alpha_{t+1}$ via Eq.~\eqref{eq:alpha_update}
    \State \textbf{if} $t \geq T_{\min}$ \textbf{and} $\mathrm{FOM}^{(t)} < \varepsilon$ \textbf{then break}
\EndFor
\State \Return $\hat{P}_f = \frac{1}{|\mathcal{D}|} \displaystyle\sum_{i} I(\mathbf{x}_i) \frac{p(\mathbf{x}_i)}{h_{t_i}(\mathbf{x}_i)}$
\end{algorithmic}
\end{algorithm}

\section{Experimental Results}
\label{sec:exp}

\subsection{Experimental Setup}

\noindent\textbf{Benchmark.}
All experiments are conducted on OpenYield~\cite{openyield}, an open-source SRAM yield analysis platform built on FreePDK45 (45\,nm) technology.
Unlike simplified bit-cell models, OpenYield provides macro-level SRAM designs with full peripheral circuits, capturing second-order effects including parasitics, inter-cell leakage, and peripheral circuit variations.
We evaluate read delay as the primary performance metric; a failure is declared when read delay exceeds a predefined timing constraint.
Process variations are modeled as independent Gaussian distributions with standard deviation set to 5\% of nominal values.
We evaluate $3\times2$, $8\times4$, $7\times6$, $8\times6$, and $8\times8$ arrays with $D=108$, $576$, $756$, $864$, and $1152$, respectively, all at 1.0\,V and TT/25$^\circ$C.

\noindent\textbf{Baselines.}
We compare against eight importance sampling baselines: MNIS~\cite{mnis}, HSCS~\cite{hscs}, AIS~\cite{ais}, ACS~\cite{acs}, OPT~\cite{opt}, VIS~\cite{vis}, FUSIS~\cite{fusis}, and PCA+MNIS.
Monte Carlo (MC) with 61,000--100,000 samples serves as the reference estimator.
All methods are evaluated in the same OpenYield environment with identical initial sampling budgets and SPICE simulation settings.
For fairness in rare-event regimes, baseline hyperparameters are tuned following established practices.
When non-convergence is observed, typically indicated by persistently large relative errors or weight degeneracy, sampling budgets are increased by enlarging the initial sample size or raising the per-iteration allocation to maintain a minimum number of effective failure samples.
Entries marked NA indicate methods that failed to converge across all attempted runs at that configuration.

\noindent\textbf{Hyperparameters.}
The TabPFN context budget is $C_{\max}=1024$, and the SVD proposal uses the top $m=50$ samples. All other settings are shared across configurations except the target hit rate $\tau$, which decreases with the observed failure rate from 0.20 at $D=108$ to 0.08 at $D=1152$. We separately test a fixed $\tau=0.1$ below to assess whether this schedule is necessary.

\begin{table*}[!t]
\begin{minipage}[t]{0.485\textwidth}
\centering
\footnotesize
\caption{End-to-end runtime at $D=576$ (seconds). SPICE time is normalized at 5.25 seconds per evaluation.}
\label{tab:runtime}
\setlength{\tabcolsep}{2pt}
\renewcommand{\arraystretch}{0.82}
\begin{tabular*}{\linewidth}{@{\extracolsep{\fill}}lrrrr@{}}
\toprule
\textbf{Method} & \textbf{Model} & \textbf{SPICE} & \textbf{Total} & \textbf{Speedup} \\
\midrule
MC     & 0.0      & 525,575 & 525,575 & $1.0\times$ \\
HOLMES & 637.1    & 26,542  & 27,179  & $19.3\times$ \\
OPT    & 1,015.6  & 48,800  & 49,815  & $10.6\times$ \\
VIS    & 550.8    & 76,208  & 76,759  & $6.9\times$ \\
FUSIS  & 70,944.2 & 32,954  & 103,898 & $5.1\times$ \\
\bottomrule
\end{tabular*}
\end{minipage}\hfill
\begin{minipage}[t]{0.485\textwidth}
\centering
\footnotesize
\caption{Statistical robustness over five independent seeds with fixed $\tau=0.1$.}
\label{tab:multiseed}
\setlength{\tabcolsep}{2pt}
\renewcommand{\arraystretch}{0.82}
\begin{tabular*}{\linewidth}{@{\extracolsep{\fill}}cccc@{}}
\toprule
$D$ & \textbf{Ref. $P_f$} & \textbf{$P_f$ Estimate ($\mu\pm\sigma$)} & \textbf{CV} \\
\midrule
108 & 3.61\%  & $3.73\%\pm0.09\%$   & 2.5\% \\
576 & 1.142\% & $1.162\%\pm0.042\%$ & 3.6\% \\
756 & 0.879\% & $0.896\%\pm0.035\%$ & 3.9\% \\
864 & 0.736\% & $0.731\%\pm0.021\%$ & 2.9\% \\
1152 & 0.2629\% & $0.269\%\pm0.011\%$ & 4.0\% \\
\bottomrule
\end{tabular*}
\end{minipage}
\end{table*}

\subsection{Main Results}
\label{sec:main_results}
\noindent\textbf{Computational overhead.}
At $D=576$, model computation accounts for only 637.1 seconds (2.3\%) of HOLMES's total runtime, 37\% less than OPT (Table~\ref{tab:runtime}). The resulting $19.3\times$ wall-clock speedup is close to the $19.8\times$ simulation-count speedup in Table~\ref{tab:main_results}, showing that TabPFN inference does not offset the saved SPICE evaluations.

\noindent\textbf{Fixed-$\tau$ robustness.}
Across five independent runs, fixed $\tau=0.1$ yields mean relative errors of 3.3\%, 1.8\%, and 0.7\% at $D=108$, 576, and 864, respectively, with CV at most 3.6\% (Table~\ref{tab:multiseed}). Thus, dimension-specific scheduling is not required for stable estimates on the three tested configurations.

\noindent\textbf{Dimension-wise comparison.} Table~\ref{tab:main_results} summarizes the central finding: every full-dimension baseline collapses at some dimension, with peak errors of 24.5--100\%, whereas HOLMES remains within 5.9\%. The table also adds head-to-head VIS, FUSIS, and PCA+MNIS results. Some baselines exhibit \emph{false convergence}, stabilizing around an incorrect estimate without a warning from the FOM criterion.
\begin{table*}[!t]
\centering
\scriptsize
\caption{Unified yield-estimation results across all dimensions. Rel. Err. reports absolute relative error; PCA+MNIS was evaluated only at $D=864$ and $D=1152$. Panels are arranged in a three-plus-two layout.}
\label{tab:main_results}
\setlength{\tabcolsep}{1.2pt}
\setlength{\extrarowheight}{0pt}
\setlength{\aboverulesep}{0.8pt}
\setlength{\belowrulesep}{0.8pt}
\renewcommand{\arraystretch}{0.78}
\newcommand{\resultpanel}[2]{%
\begin{minipage}[t]{0.32\textwidth}\centering
\fontsize{5.8pt}{6.4pt}\selectfont
\begin{tabular*}{\linewidth}{@{\extracolsep{\fill}}lrrrr@{}}
\toprule
\multicolumn{5}{c}{\textbf{$D=#1$}}\\
\textbf{Method} & \textbf{Est. Fail Rate} & \textbf{Rel. Err. (\%)} & \textbf{Sim.} & \textbf{Speedup}\\
\midrule
#2
\bottomrule
\end{tabular*}
\end{minipage}%
}
\resultpanel{108}{%
MC       & $3.612\times10^{-2}$ & --   & 100,000 & $1.0\times$ \\
ACS      & $4.848\times10^{-2}$ & 34.2 & 4,100   & $24.4\times$ \\
HSCS     & $4.041\times10^{-2}$ & 11.9 & 7,500   & $13.3\times$ \\
MNIS     & $3.899\times10^{-2}$ & 7.9  & 2,700   & $37.0\times$ \\
AIS      & $3.125\times10^{-2}$ & 13.5 & 6,800   & $14.7\times$ \\
OPT      & $3.896\times10^{-2}$ & 7.9  & 5,322   & $18.8\times$ \\
VIS      & $3.848\times10^{-2}$ & 6.5  & 6,233   & $16.1\times$ \\
FUSIS (Avg.) & $1.410\times10^{-3}$ & 96.1 & 5,970 & $18.1\times$ \\
\textbf{HOLMES} & $\mathbf{3.624\times10^{-2}}$ & \textbf{0.34} & \textbf{1,700} & $\mathbf{58.8\times}$ \\
}\hfill
\resultpanel{576}{%
MC       & $1.142\times10^{-2}$ & --   & 100,000 & $1.0\times$ \\
ACS      & $8.333\times10^{-3}$ & 27.0 & 8,800   & $11.4\times$ \\
HSCS     & $1.363\times10^{-2}$ & 19.4 & 8,200   & $12.2\times$ \\
MNIS     & $1.081\times10^{-2}$ & 5.3  & 9,500   & $10.5\times$ \\
AIS      & $1.190\times10^{-2}$ & 4.2  & 10,080  & $9.9\times$ \\
OPT      & $1.188\times10^{-2}$ & 4.0  & 9,285   & $10.8\times$ \\
VIS      & $1.549\times10^{-2}$ & 35.7 & 14,500  & $7.0\times$ \\
FUSIS    & $4.400\times10^{-4}$ & 96.1 & 6,173   & $16.4\times$ \\
\textbf{HOLMES} & $\mathbf{1.156\times10^{-2}}$ & \textbf{1.2} & \textbf{5,050} & $\mathbf{19.8\times}$ \\
}\hfill
\resultpanel{756}{%
MC       & $8.790\times10^{-3}$ & --   & 70,200 & $1.0\times$ \\
ACS      & $6.758\times10^{-3}$ & 23.1 & 6,100  & $11.5\times$ \\
HSCS     & $1.272\times10^{-2}$ & 44.7 & 8,800  & $8.0\times$ \\
MNIS     & $9.259\times10^{-3}$ & 5.3  & 7,200  & $9.8\times$ \\
AIS      & $7.523\times10^{-3}$ & 14.4 & 14,160 & $5.0\times$ \\
OPT      & $1.130\times10^{-5}$ & 99.9 & 19,183 & $3.7\times$ \\
VIS      & $7.690\times10^{-3}$ & 12.6 & 17,800 & $3.9\times$ \\
FUSIS    & $2.300\times10^{-4}$ & 97.4 & 7,000  & $10.0\times$ \\
\textbf{HOLMES} & $\mathbf{8.554\times10^{-3}}$ & \textbf{2.7} & \textbf{3,300} & $\mathbf{21.3\times}$ \\
}\par\vspace{2pt}
\makebox[\textwidth][c]{%
\resultpanel{864}{%
MC       & $7.360\times10^{-3}$  & --    & 61,000 & $1.0\times$ \\
ACS      & $7.317\times10^{-3}$  & 0.6   & 7,400  & $8.2\times$ \\
HSCS     & $8.592\times10^{-3}$  & 16.7  & 7,800  & $7.8\times$ \\
MNIS     & $5.555\times10^{-3}$  & 24.5  & 9,800  & $6.2\times$ \\
AIS      & $7.812\times10^{-3}$  & 6.1   & 12,360 & $4.9\times$ \\
OPT      & $3.780\times10^{-83}$ & 100.0 & 19,927 & $3.1\times$ \\
VIS      & $6.880\times10^{-3}$  & 6.5   & 20,233 & $3.0\times$ \\
FUSIS    & $3.500\times10^{-4}$  & 95.23 & 12,540 & $4.9\times$ \\
PCA+MNIS & $7.780\times10^{-3}$  & 5.67  & 7,000  & $8.7\times$ \\
\textbf{HOLMES} & $\mathbf{7.353\times10^{-3}}$ & \textbf{0.1} & \textbf{5,600} & $\mathbf{10.9\times}$ \\
}\hspace{0.02\textwidth}
\resultpanel{1152}{%
MC       & $2.629\times10^{-3}$   & --    & 100,000 & $1.0\times$ \\
ACS      & $2.941\times10^{-3}$   & 11.9  & 38,600  & $2.6\times$ \\
HSCS     & $1.351\times10^{-3}$   & 48.6  & 28,300  & $3.5\times$ \\
MNIS     & $1.531\times10^{-3}$   & 67.8  & 40,800  & $2.5\times$ \\
AIS      & $3.307\times10^{-3}$   & 25.8  & 34,920  & $2.9\times$ \\
OPT      & $8.630\times10^{-213}$ & 100.0 & 25,567  & $3.9\times$ \\
VIS      & $2.089\times10^{-3}$   & 20.5  & 41,800  & $2.4\times$ \\
FUSIS    & $1.310\times10^{-4}$   & 95.0  & 17,561  & $6.3\times$ \\
PCA+MNIS & $2.298\times10^{-3}$   & 12.50 & 17,300  & $5.78\times$ \\
\textbf{HOLMES} & $\mathbf{2.785\times10^{-3}}$ & \textbf{5.9} & \textbf{13,000} & $\mathbf{7.7\times}$ \\
}%
}
\end{table*}
HOLMES uses 1,700--13,000 simulations and is the only method bounded by 5.9\% error across all dimensions. OPT falsely converges with 99.9\% error at $D=756$ and becomes numerically degenerate thereafter; MNIS reaches 24.5\% and 67.8\% error at $D=864$ and $1152$. Dimension reduction helps but does not close the gap: PCA+MNIS records 5.67\% and 12.50\% error at these dimensions, versus 0.1\% and 5.9\% for HOLMES.

\begin{figure}[!t]
    \centering
    \includegraphics[width=\linewidth]{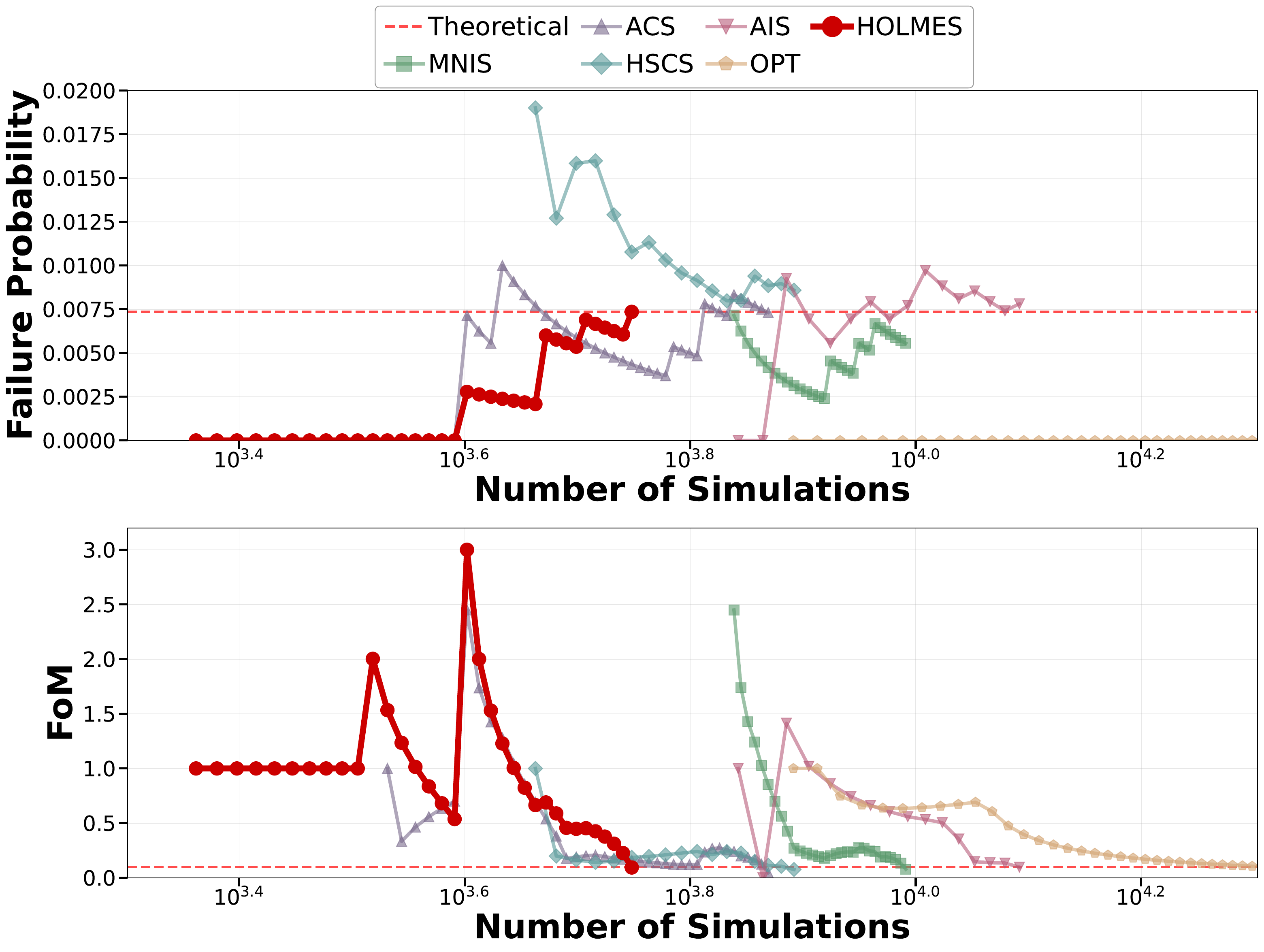}
    \caption{Convergence at $D = 864$.}
    \label{fig:conv_864}
\end{figure}

\begin{figure}[!t]
    \centering
    \includegraphics[width=\linewidth]{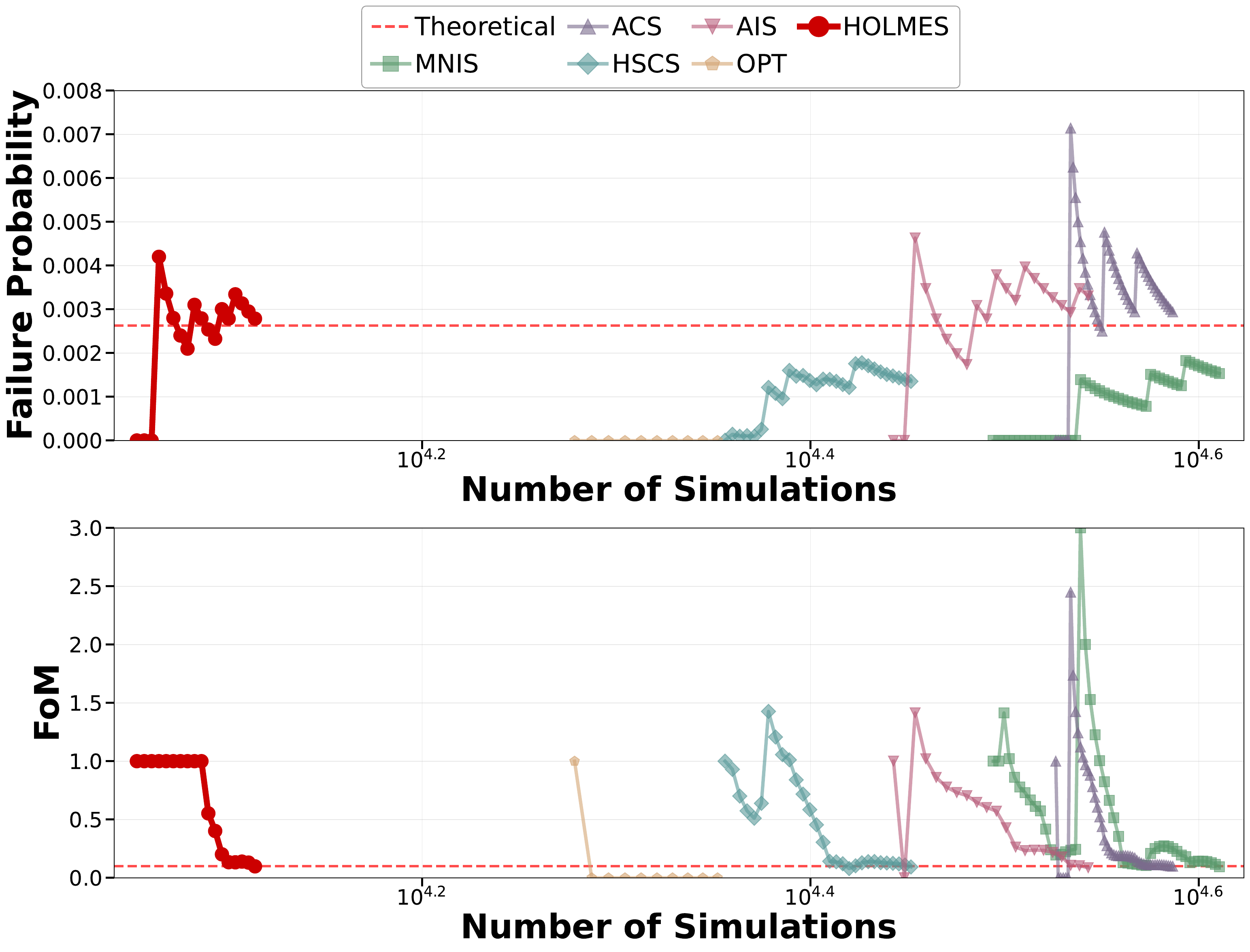}
    \caption{Convergence at $D = 1152$.}
    \label{fig:conv_1152}
\end{figure}

\subsection{Classifier Substitution Analysis}

At $D=864$, we replace only TabPFN while holding all sampling components fixed. Table~\ref{tab:rel_err} shows 28.7--73.8\% error for conventional classifiers versus 0.1\% for TabPFN, isolating failure-center localization as the decisive factor under extreme imbalance.

\begin{table}[!t]
\centering
\scriptsize
\caption{Classifier substitution at $D = 864$ (MC reference: $7.36 \times 10^{-3}$). All variants use identical proposal and mixing components; only the localization classifier differs.}
\label{tab:rel_err}
\setlength{\tabcolsep}{2pt}
\setlength{\aboverulesep}{1.5pt}
\setlength{\belowrulesep}{1.5pt}
\renewcommand{\arraystretch}{0.80}
\begin{tabular*}{\linewidth}{@{\extracolsep{\fill}}lccc}
\toprule
\textbf{Classifier} & \textbf{Est. Fail Rate} & \textbf{Sim} & \textbf{Rel. Err. (\%)} \\
\midrule
MC                  & $7.36 \times 10^{-3}$ & 61,000 & 0.0 \\
RF                  & $5.25 \times 10^{-3}$ & 8,000   & 28.7 \\
QDA                 & $4.97 \times 10^{-3}$ & 6,900   & 32.5 \\
NB                  & $4.32 \times 10^{-3}$ & 8,000   & 41.3 \\
GP                  & $4.14 \times 10^{-3}$ & 8,800   & 43.8 \\
GBDT                & $3.74 \times 10^{-3}$ & 9,300   & 49.2 \\
SVC                 & $9.49 \times 10^{-3}$ & 7,800   & 28.9 \\
DT                  & $3.11 \times 10^{-3}$ & 8,000   & 57.7 \\
LR                  & $2.02 \times 10^{-3}$ & 8,600   & 72.6 \\
MLP                 & $1.05 \times 10^{-2}$ & 6,800   & 42.6 \\
AdaBoost            & $1.96 \times 10^{-3}$ & 8,200   & 73.8 \\
\textbf{TabPFN}     & $\mathbf{7.35 \times 10^{-3}}$ & \textbf{5,600} & \textbf{0.1} \\
\bottomrule
\end{tabular*}
\end{table}

\subsection{Ablation Study}

Table~\ref{tab:ablation} removes each HOLMES component at three representative dimensions.

\begin{table}[!t]
\centering
\scriptsize
\caption{Ablation study. MC serves as the reference at each dimension.}
\label{tab:ablation}
\setlength{\tabcolsep}{2pt}
\setlength{\aboverulesep}{1.5pt}
\setlength{\belowrulesep}{1.5pt}
\renewcommand{\arraystretch}{0.80}
\begin{tabular*}{\linewidth}{@{\extracolsep{\fill}}clcc@{}}
\toprule
\textbf{Dim} & \textbf{Variant} & \textbf{Est. Fail Rate} & \textbf{Rel Err} \\
\midrule
\multirow{4}{*}{108}
& True Rate             & 3.612e-2 & 0\%     \\
& \textbf{HOLMES}       & 3.6242e-2 & 0.34\%  \\
& \quad w/o SVD         & 4.137e-2 & 14.53\% \\
& \quad w/o Adapt       & 5.091e-2 & 40.95\% \\
\midrule
\multirow{4}{*}{576}
& True Rate             & 1.142e-2 & 0\%     \\
& \textbf{HOLMES}       & 1.156e-2 & 1.26\%  \\
& \quad w/o SVD         & 9.094e-3 & 20.36\% \\
& \quad w/o Adapt       & 1.000e-2 & 12.43\% \\
\midrule
\multirow{4}{*}{864}
& True Rate             & 7.360e-3 & 0\%     \\
& \textbf{HOLMES}       & 7.353e-3 & 0.10\%  \\
& \quad w/o SVD         & 8.2347e-3 & 11.88\% \\
& \quad w/o Adapt       & 6.944e-3 & 5.65\%  \\
\bottomrule
\end{tabular*}
\end{table}

Removing the anisotropic proposal causes a substantial accuracy loss at every tested dimension: relative error is 14.53\%, 20.36\%, and 11.88\% at $D=108$, 576, and 864, respectively. Removing adaptive mixing is most damaging at $D=108$ (40.95\%), but remains less disruptive at the higher dimensions tested. These results show that SVD alignment and hit-rate adaptation provide complementary safeguards for proposal quality and exploration.

\subsection{Failure Mode Analysis}
\label{sec:failure_modes}

Table~\ref{tab:main_results} and Figures~\ref{fig:conv_864}--\ref{fig:conv_1152} reveal two distinct failure modes. In \emph{false convergence}, the FOM becomes stable even though the estimate is wrong: OPT at $D=756$ reaches 99.9\% relative error, and MNIS at $D=864$ reaches 24.5\% error while satisfying the stopping criterion. The FOM measures short-term estimator stability, not proximity to the true failure probability; once a method settles around an incorrectly localized center, it can therefore return a confident but invalid estimate. In \emph{non-convergence}, the estimated center keeps shifting as new failures arrive, producing oscillating FOM trajectories or numerically degenerate weights at the largest dimensions. These two behaviors have the same underlying cause: sparse failures make gradient-based or geometry-only localization unreliable in the high-dimensional space. HOLMES avoids both modes across all five configurations: TabPFN supplies a failure-preserving, gradient-free ranking, while the SVD proposal and hit-rate mixing retain exploration until the estimate stabilizes near the correct value.

\section{Conclusion}

Surrogate-assisted importance sampling methods for high-sigma yield estimation share a structural assumption that has gone unexamined: that locating the failure center requires fitting a parametric model to an inherently imbalanced dataset.
We show this assumption is unnecessary.
Recasting failure-center localization as few-shot binary classification enables gradient-free in-context inference via a prior-fitted tabular foundation model, eliminating the ill-posed training loop without sacrificing localization accuracy.

{\setlength{\emergencystretch}{1em}%
HOLMES combines this reformulation with an SVD-based 
aniso\-tropic proposal and a hit-rate-driven adaptive mixing scheme.
The ablation study confirms that both components are necessary and complementary: removing the SVD proposal increases error from 0.1\% to 11.88\% at $D = 864$, while removing adaptive mixing increases error to 40.95\% at $D = 108$.
On 6T SRAM benchmarks spanning $D = 108$ to $D = 1{,}152$, HOLMES remains within 5.9\% relative error across all five configurations; full-dimensional baselines exhibit accuracy collapse at some dimension, with the strongest baseline reaching 25.8\% error, while PCA+MNIS is evaluated at the two largest dimensions.
The current formulation assumes a single dominant failure region; extending the framework to multi-center localization for circuits with well-separated failure modes is a natural direction for future work.
\par}

\bibliographystyle{ACM-Reference-Format}
\bibliography{refs}

\end{document}